\documentclass[prl,twocolumn,showpacs,superscriptaddress,floatfix]{revtex4}

\usepackage{graphicx}
\usepackage{array}
\usepackage{amsmath,amsfonts,amssymb}

\begin{document}

\title{Surface State Stark Shift in a Scanning Tunneling Microscope}

\author{L. Limot}
\affiliation{Institut f\"{u}r Experimentelle und Angewandte Physik, Christian-Albrechts-Universit\"{a}t zu Kiel, D-24098 Kiel, Germany}
\author{T. Maroutian}
\affiliation{Institut f\"{u}r Experimentelle und Angewandte Physik, Christian-Albrechts-Universit\"{a}t zu Kiel, D-24098 Kiel, Germany}
\author{P. Johansson}
\affiliation{Department of Natural Sciences, University of \"{O}rebro, S-70182, \"{O}rebro, Sweden}
\author{R. Berndt}
\affiliation{Institut f\"{u}r Experimentelle und Angewandte Physik, Christian-Albrechts-Universit\"{a}t zu Kiel, D-24098 Kiel, Germany}

\begin{abstract}
We report a quantitative low-temperature Scanning Tunneling Spectroscopy (STS)
study on the Ag(111) surface state over an unprecedented range of currents ($50$
pA to $6$ $\mu$A) through which we can tune the electric field in the tunnel
junction of the microscope. We show that in STS a sizeable Stark effect causes a
shift of the surface state binding energy $E_{0}$. Data taken are reproduced by a
one-dimensional potential model calculation, and are found to yield a Stark-free
energy $E_{0}$ in agreement with recent state-of-the-art Photoemission
Spectroscopy measurements.
\end{abstract}


\maketitle

Surface states have been investigated intensely for over two decades of surface
science. These states, which are trapped between the surface barrier potential and
a band gap in the crystal, are an experimental realization of a
quasi-two-dimensional electron gas with a characteristic dispersion of
$E(k)-E_{0}\propto k^{2}$, where $E_{0}$ is the lower edge of the energy band.
Scanning Tunneling Microscopy (STM) and Spectroscopy (STS) have provided a new way
to study surface states through local measurements performed on the atomic scale.
The first imaging of Shockley surface states on the (111) facet of the noble
metals Ag, Cu and Au \cite{Crommie Hasegawa}, has motivated a great deal of STM
and STS studies on these systems. Since then, many aspects have been elucidated
such as surface-state-mediated interactions \cite{ssinteractions}, and lateral
confinement to nano-cavities \cite{ssconfinement}. The lifetime $\tau$ of these
states at $E_{0}$, as determined by STS \cite{Joerg,LifetimeB,LifetimeK}, also
agree remarkably well with recent state-of-the-art angle-resolved Photoemission
Spectroscopy (PES) measurements \cite{Reinert}. Despite this success, a worrying
discrepancy between STS and PES still subsists in noble metals concerning $E_{0}$,
as $E_{0}$(STS) is found to lie lower than $E_{0}$(PES) typically by $-5$ to $-20$
meV \cite{Joerg,Reinert}.

This last point is still an open question. A possible explanation would be that
unlike STS, where the local character of the measurement allows to access a
defect-free region of the surface, a PES measurement is defect-sensitive as it
integrates over a large sample area of about 1 mm$^{2}$. And defects should, in
principle, shift $E_{0}$(PES) upwards in energy \cite{PESDefects}. However defects
also broaden the PES line width ($\Gamma$(PES)$=\hbar /\tau$), and the recent
agreement found for $\tau$ between STS and PES appears to invalidate this
scenario.

In this Letter, we present STS measurements on the surface state of Ag(111) which
elucidate this discrepancy. The presence of an electric field between the STM tip
and the sample surface is known to produce a Stark shift in the STS spectra of
field emission resonance states \cite{Binnig}, or a more dramatic effect such as
band bending in semiconductors \cite{BandBending}. With this in mind, we have
tracked down the influence of the electric field on the STS spectrum of the
Ag(111) surface state. The tunneling current $I$ in a STM junction can be written
as
\begin{equation}
I \propto \text{exp}(-1.025\sqrt{\phi}\ d)
\label{tunnel}
\end{equation}
where $\phi$ is the apparent barrier height and $d$ the tip-surface distance,
hence the electric field ($\sim 1/d$) can be conveniently tuned through $I$. By
spanning $I$ over $50$ pA$-$$6$ $\mu$A and recording concomitantly the surface
state spectrum, we establish on an experimental basis the existence of a downward
Stark shift in $E_{0}$(STS). We also provide a one-dimensional model which
describes the Stark contribution. We extract the Stark-shift-free energy $E_{0}$
of Ag(111) and find it to agree with the PES value (where no electric field is
present).

The Ag(111) surface was cleaned by Ar$^{+}$ sputter/anneal cycles, and the
measurements were performed in a custom-built ultrahigh vacuum STM operating at a
temperature of $4.6$ K, using electrochemically etched W tips, further treated
in-situ by indentations into the surface. Spectroscopy of the differential
conductance $dI/dV$ versus the sample voltage $V$ was performed by opening the
feedback loop at $V=100$ mV in the center of impurity- and step-free regions of
the surface ($\geq 400$ nm$^{2}$). A lock-in detection amplifier was employed to
record the $dI/dV$ (AC voltage modulation was $1$ mV rms and $\sim 1-10$ kHz,
and a single spectrum was acquired in $\sim 3-10$ s). The dependency of the $dI/dV$
spectrum on $d$ is probed through Eq.\ \ref{tunnel} by varying $I$.

\begin{figure}[ht]
\includegraphics[width=6.0cm,bbllx=70,bblly=265,bburx=495,bbury=780,clip=]{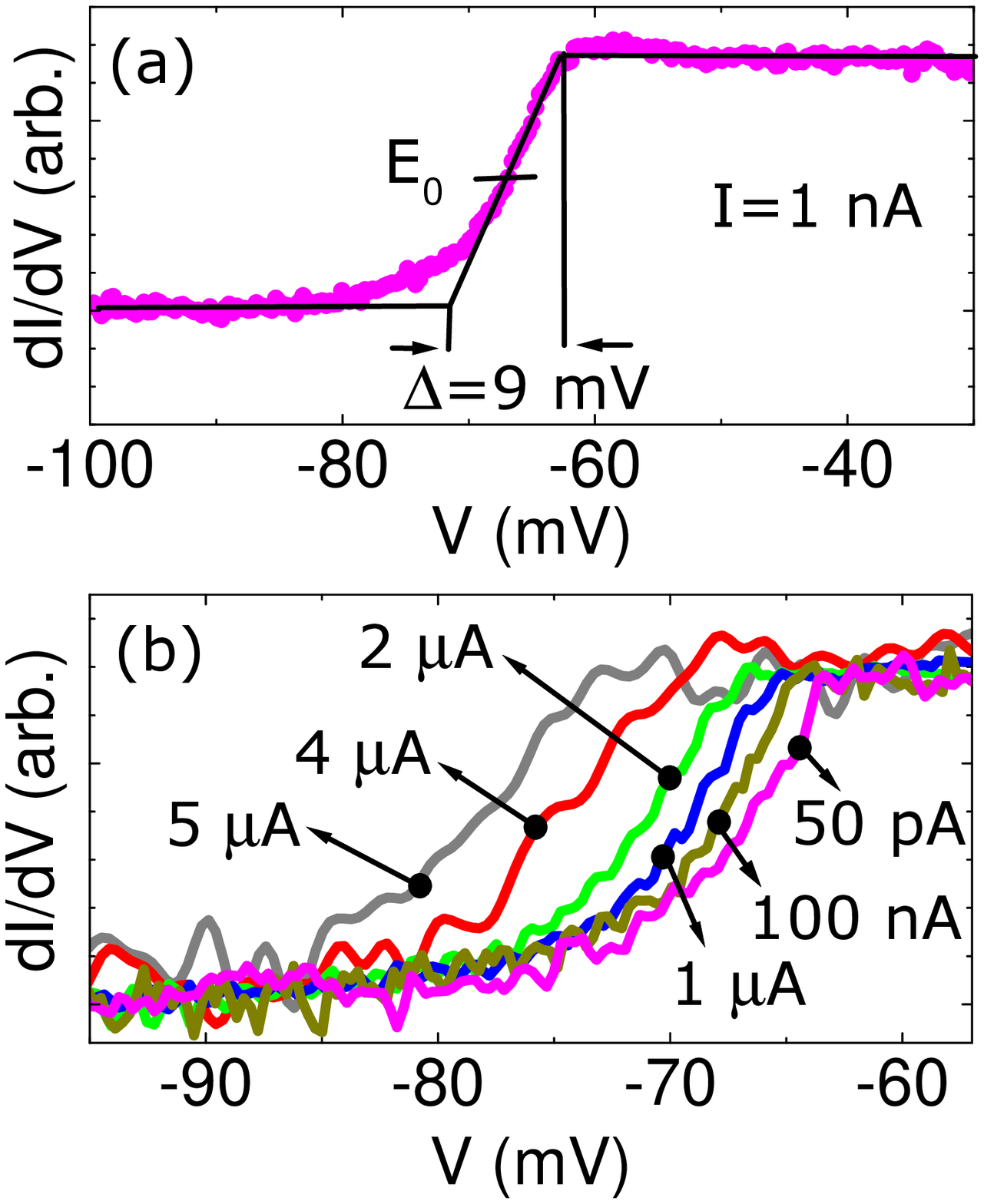}
\caption{(a) Typical $dI/dV$ spectrum taken on Ag(111) ($T=4.6$ K). The feedback
loop was opened at $I=1$ nA and $V=100$ mV. A geometrical analysis is used to
determine $E_{0}$ and the width $\Delta$ of the surface state
\cite{Joerg,LifetimeB}. (b) Downward shift of $E_{0}$ when $I$ is increased. All
spectra are averages of at least $5$ single spectra from varying sample locations
and tips; the spectra were renormalized to match the conductance of the 50 pA
spectrum.}
\label{fig1}
\end{figure}

Figure~\ref{fig1} summarizes the essential experimental findings. The sharp
step-like rise of the conductance shown in Fig.~\ref{fig1}a for $I=1$ nA allows to
determine the low edge of the energy band of the Ag(111) surface state $-$ at
$E_{0}=-66.7(5)$ meV in Fig.~\ref{fig1}a, in agreement with previous STS studies
\cite{Joerg}, but lower than $E_{0}$(PES)$=-63(1)$ meV \cite{Reinert}. In
contrast, from the width $\Delta$ of the onset we extract $\Gamma$(STS)$=6.4(5)$
meV \cite{LifetimeB}, in agreement with $\Gamma$(PES)$=6.0(5)$. When increasing
$I$, a sizeable shift of $E_{0}$ occurs. In typical $dI/dV$ spectra
(Fig.~\ref{fig1}b) the surface state onset shifts downward in energy with no
appreciable broadening as $I$ increases from a pA-range to a $\mu$A-range. At
$I=6$ $\mu$A, the highest current where we could perform reliable spectroscopy,
$E_{0}$ has shifted by $-20\%$ with respect to $I=50$ pA. The dependence of
$E_{0}$ on $I$, i.e. on the tip-surface distance, is the major finding of this
Letter.

\begin{figure}[t]
\includegraphics[width=6.0cm,bbllx=60,bblly=415,bburx=530,bbury=745,clip=]{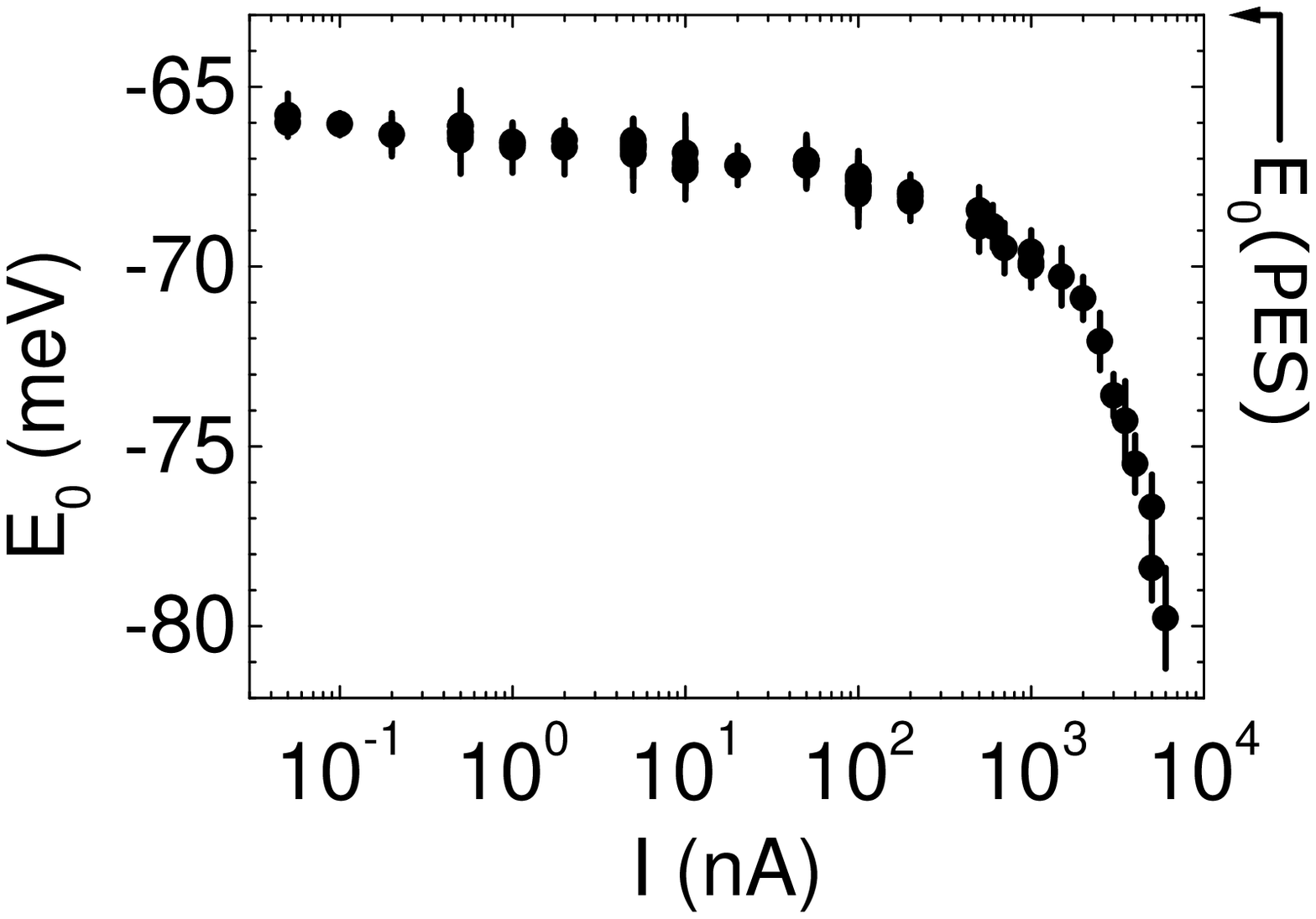}
\caption{$E_{0}$ versus $I$. The arrow indicates the PES value of $E_{0}$ \cite{Reinert}.}
\label{fig2}
\end{figure}

Figure~\ref{fig2} presents a quantitative evaluation of $E_{0}$ for all tunneling
currents investigated. Two distinct regimes are found for the band edge shift $-$
also directly visible on the spectra of Fig.~\ref{fig1}b: typically $50$ pA$\leq
I\leq 1$ $\mu$A over which $E_{0}$ decreases by $-4$ meV, and $2$ $\mu$A$<I\leq 6$
$\mu$A where the shift of $E_{0}$ is more pronounced ($-10$ meV).

Care was taken to systematically survey the tip and the surface status during
spectrum acquisition. We proceeded as follows: (i) for a given tip preparation a
$I=1$ nA spectrum is acquired and $E_{0}$ is evaluated following the geometrical
analysis detailed in Fig.~\ref{fig1}a, (ii) $I$ is then set to the value of
interest and a spectrum is recorded, (iii) the tip and the surface status are
checked through a final $1$ nA spectrum. To minimize errors in the evaluation of
the shift, each spectrum acquired in step (ii) was fitted to the 1 nA spectrum of
step (i) in order to determine $E_{0}$. For all the data reported in this Letter,
no change was discernable in the spectra acquired in step (i) and in step (iii),
meaning that neither a tip modification nor a tip-induced damage of the surface
occurred when acquiring spectra in the $50$ pA $-$ $6$ $\mu$A range. This
conclusion also agrees with the absence of any modification in the surface images
acquired before step (i) and after step (iii).

\begin{figure}[b]
\includegraphics[width=6.0cm,bbllx=50,bblly=245,bburx=500,bbury=730,clip=]{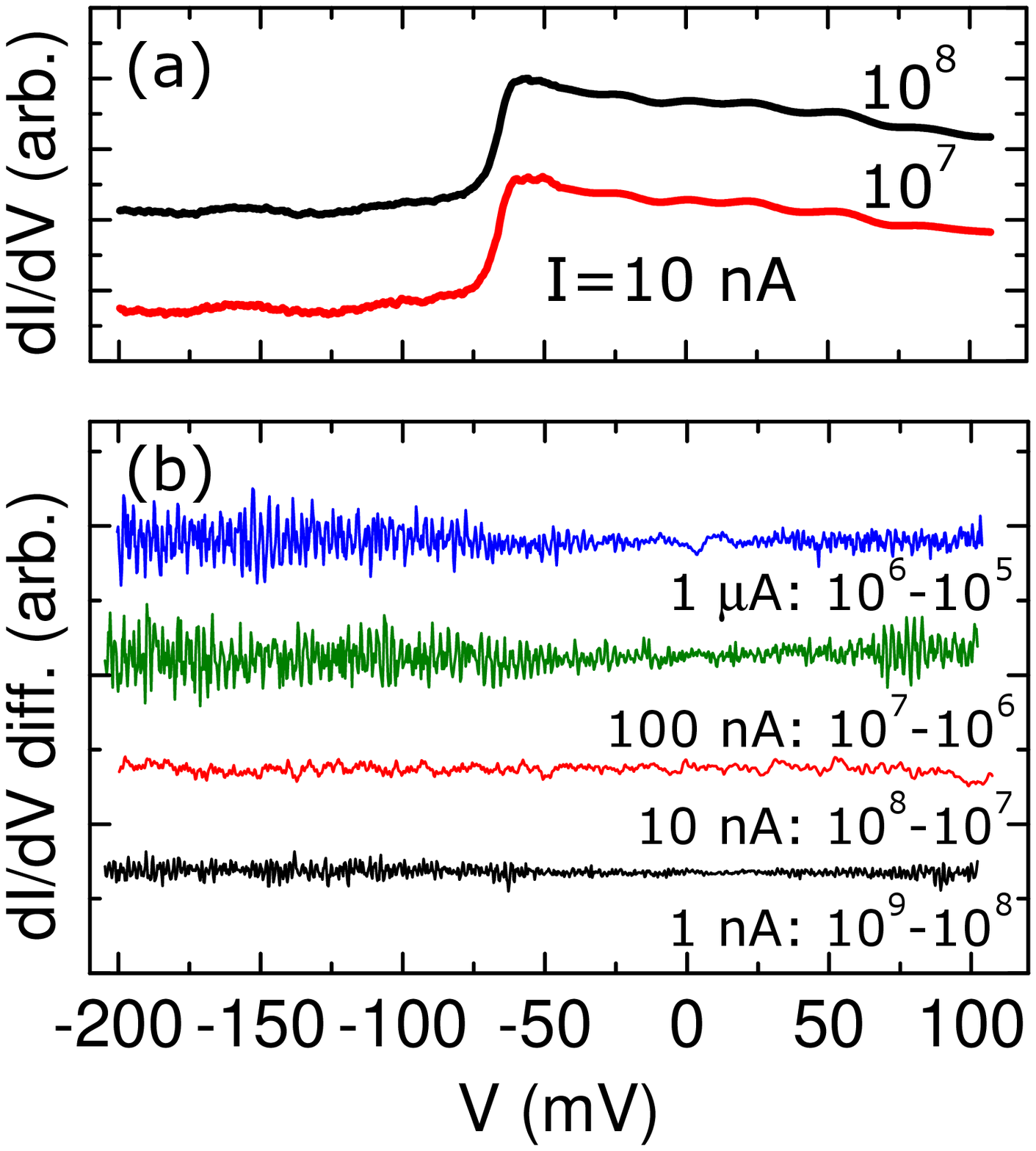}
\caption{(a) $dI/dV$ spectra acquired at $10$ nA, $100$ mV,  with a $10^{7}$ and a
$10^{8}$ V/A gain of the current preamplifier. (b) Differences between pairs of
spectra acquired at the same currents (1 nA, 10 nA, 100 nA and 1 $\mu$A) but with
different gains. Gains range from $10^{5}$ to $10^{9}$ V/A with $R_{in}=50$
$\Omega$, $60$ $\Omega$, $150$ $\Omega$, $1$ k$\Omega$ and $10$ k$\Omega$,
respectively. Spectra are shifted vertically for clarity.}
\label{fig3}
\end{figure}

Before going any further, a remark concerning the experimental setup is necessary.
Given the unusually wide range of the tunneling resistances $R$ employed in this
study (from $20$ k$\Omega$ to $2$ G$\Omega$), the possibility of a voltage drop at
the input impedance $R_{in}$ of the current preamplifier has to be considered. It
would decrease the bias voltage at the tunneling junction by $(1+R_{in}/R)^{-1}$
thus producing a sizeable $R$-dependent, i.e. $I$-dependent, shift of the spectrum
when $R\sim R_{in}$. With the variable gain current preamplifier employed
\textit{this does not occur}, since $R_{in}/R\approx 0.3\%$ at most
(this condition is met for both DC and AC signals at the modulation frequencies used).
To highlight this point, Fig.~\ref{fig3}a displays
two spectra acquired at $10$ nA but with gains of $10^{7}$ and $10^{8}$ V/A for which
$R_{in}/R=0.02\%$ and $R_{in}/R=0.1\%$ respectively. As expected, their difference is a flat line over
the entire voltage range (Fig.~\ref{fig3}b), even in the onset region where a
shift artifact would produce a spike. This is also true for pairs of spectra
acquired at 1 nA, 100 nA and 1$\mu$A at different gains. In conclusion, no
relevant voltage drop is present in our setup at the tunneling currents of
interest.

In order to express the dependency of $E_{0}$ on the tip-surface distance, we
measured the variation of $I$ while approaching, with the feedback loop open, the
tip towards the surface by $\approx 6$ \AA\@. The recorded currents which cover
$20$ pA $-$ $6$ $\mu$A follow an exponential behavior up to $\approx 2$ $\mu$A,
from which we extract through Eq.\ \ref{tunnel} an apparent barrier height of
$\phi=4.0(2)$ eV, typical for noble metals \cite{Besenbacher}. In contrast, for
$I\agt 2$ $\mu$A, $I$ increases faster than expected. It is well known from break
junction experiments \cite{Krans}, that this signals that the junction is no
longer in a tunneling regime, rather in a contact regime where, as we discuss
below, important modifications of the tip and the surface morphologies occur.

\begin{figure}[t]
\includegraphics[width=5.9cm,bbllx=30,bblly=280,bburx=505,bbury=760,clip=]{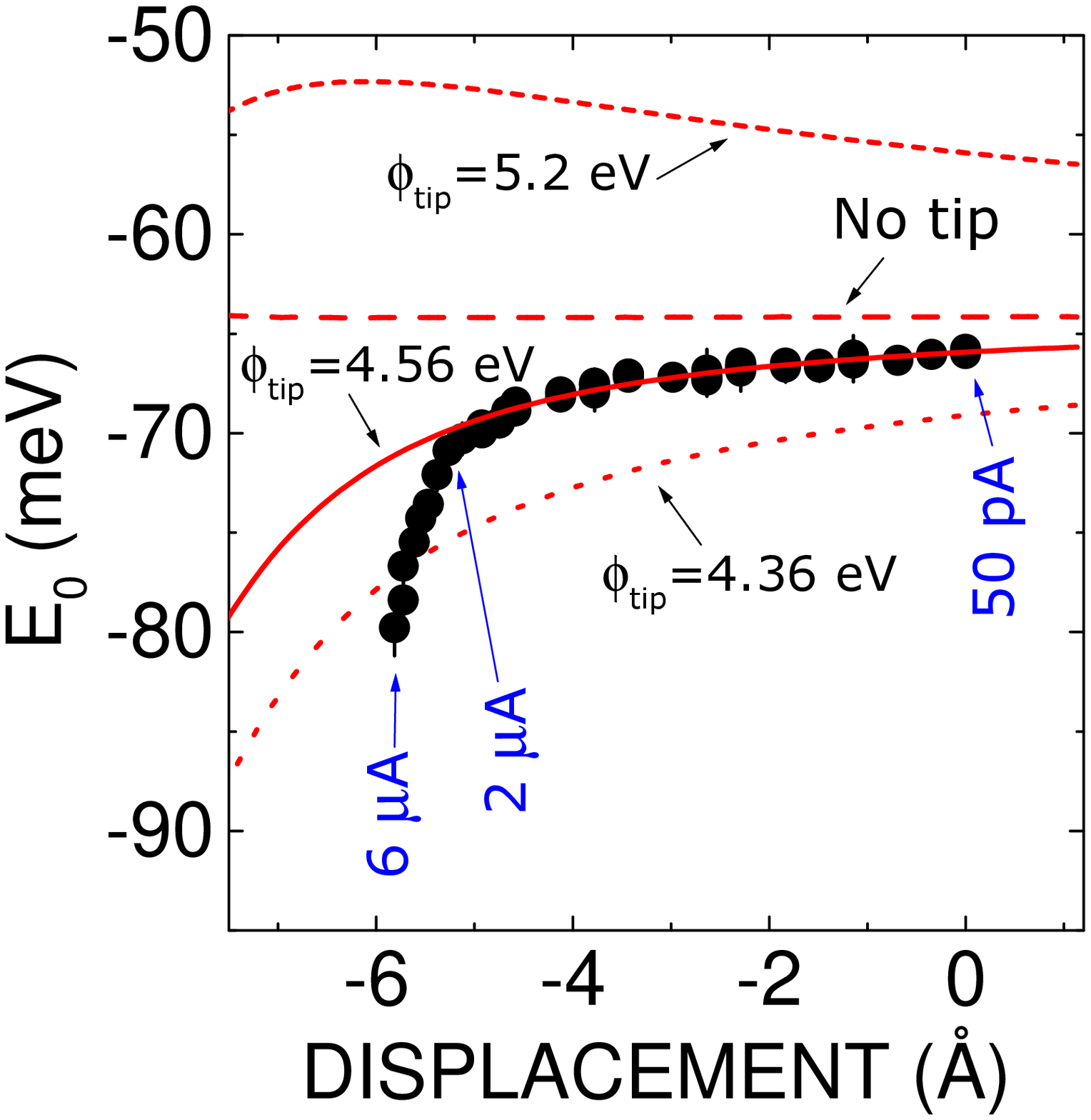}
\caption{$E_{0}$ versus tip-surface distance. Since experimentally we only probe a
displacement of the tip towards the surface, the calculated curves were shifted
horizontally to match the data (the origin is arbitrarily fixed at 50 pA). Lines
show calculated results for tip work functions $\phi_{\rm tip}$: $4.56$ eV
(solid), $4.36$ eV (dot), $5.2$ eV (short dash), and in absence of the tip (dash).
Also shown, the boundaries to the $2-6$ $\mu$A region.}
\label{fig4}
\end{figure}

Our results are summarized in Fig.~\ref{fig4}, where $E_{0}$ is now plotted versus
the tip-surface displacement. As shown, the negative Stark shift of $E_{0}$ can be
tuned by approaching or retracting the tip from the surface. To gain further
insight, we propose a model which describes the experimental data and enables
extrapolation to the zero-field properties of the surface state. We first detail
some aspects of the model and then discuss the calculation in the light of our
data.

The Ag(111) surface state electrons are modeled with the one-dimensional potential
proposed by Chulkov {\it et al.} \cite{Chulkov}. This potential is periodic in the
bulk ($z<0$), has a potential well just outside the surface ($0<z<z_1$), then
decays exponentially towards the vacuum ($z_1<z<z_{\rm im}$) to finally cross over
to a long-range image potential ($z>z_{\rm im}$). To take into account the
presence of the tip at $z\geq z_{\mathrm{tip}}$, we added to the potential the
linear contribution of the voltage bias $V$ between tip and surface, as well as
the difference between the work functions of the tip ($\phi_{\rm tip}$) and the
surface ($\phi_{\rm samp}$) to include the contact potential. Furthermore, we
modified the shape of the image potential to account for multiple images in the
tip and the surface.
This yields \begin{eqnarray}
   &&
   V(z) = 2 V_{111} \cos{gz}, \ \ z<0
   \nonumber \\ &&
   V(z) = V_{20} + V_2 \cos{\beta z}, \ \ 0<z<z_1
   \nonumber \\ &&
    V(z) = V_{\rm lin}(z)
        + V_3 e^{-\alpha(z-z_1)}, \ \ z_1<z<z_{\rm im}
   \nonumber \\ &&
    V(z) = V_{\rm lin}(z)
  - V_{\rm im}(z), \ \
   z_{\rm im}<z<z_{\rm tip}
   \nonumber
\end{eqnarray}
where
$$
   V_{\rm lin}(z) = E_{F} + s(eV+\phi_{\rm tip}) + (1-s)\phi_{\rm samp}
$$
where $E_F$ is the surface Fermi energy, $s = (z-z_1)/(z_{\rm tip}-z_1)$, and
$$
  V_{\rm im} (z) = [1 - e^{-\lambda(z-z_{\rm im})}]
   \frac{e^2\,[2\, \Psi(1) - \Psi(\zeta) - \Psi(1-\zeta)]}
   {16\pi \epsilon_0(z_{\rm im}^{\rm tip}-z_{\rm im})},
$$
with $\zeta=(z-z_{\rm im})/(z_{\rm im}^{\rm tip}-z_{\rm im})$ and $\Psi$ is the
digamma function. The parameters $V_{111}$, $g$, $V_{20}$, $V_2$, $\beta$, and
$z_1=5\pi/(4\beta)$ describing the potential in the absence of the tip are fixed
to their corresponding values of Ref.~\onlinecite{Chulkov}. The remaining
parameters $V_3$, $\alpha$, $\lambda$, and $z_{\rm im}$ and $z_{\rm im}^{\rm tip}$
that account for the tip contribution, were fixed by requiring the potential and
its derivative to be continuous everywhere, except at $z=z_{\rm tip}$ for the
derivative. The energy of the surface state at a given tip-surface distance was
calculated by searching for the corresponding maximum in the transmission
probability for electrons tunneling from the tip to the surface (a small imaginary
part was introduced in the potential between $0\le z \le z_1$ to mimic the
inelastic scattering).

The calculation performed with the Ag(111) work function $\phi_{\rm
samp}=\phi_{\rm tip}=4.56$ eV for both tip and surface
(solid line in Fig.~\ref{fig4}) reproduces the data up to $I \approx 2$ $\mu$A \cite{EF}, where a strong
discrepancy is observed. We first focus on the $I \alt 2$ $\mu$A region were the
model works. To probe the influence of the contact potential, we performed
calculations with different values for $\phi_{\rm tip}$: for a pure W tip
($\phi_{\rm tip}=5.20$ eV, short dashed line on Fig.~\ref{fig4}), and for a Ag
coated tip with a high density of surface defects ($\phi_{\rm tip}=4.36$ eV,
dotted line). The contact potential yields, in the former case, a Stark shift
opposite to the one of the negative bias, whereas the latter increases the
downward shift. As shown, neither of them reproducse the experimental behavior as
well as the calculation with $\phi_{\rm tip}=4.56$ eV does. Although we cannot
fully exclude a contribution from the contact potential, from these considerations
we conclude that, for the tips used, it must bear a negligible contribution to the
Stark effect. We note also that since the data was obtained with a variety of
tips, the highly reproducible shift observed indicates that our in-situ tip
preparation must lead to similar tip apices. In
conclusion, on the basis of our model, at the largest tip-surface distances shown,
the Stark effect is mainly produced by the bias and by the lowering of the image
potential compared with the case of an isolated surface, with a 2 to 1 ratio
between the two. However, as tip-surface distance decreases the lowering of
the image potential is increasingly stronger and eventually governs the Stark
effect.

Now we discuss the asymptotic behavior of the Stark effect. As the tip is
retracted, the Stark contribution decreases towards the non-perturbed $E_{0}$
binding energy which is calculated by suppressing the presence of the tip in our
model. It yields a constant value (dashed line on Fig.~\ref{fig4}) with
tip-surface distance of $E_{0}=-64$ meV, in agreement with the Stark-free PES
value for Ag(111). Figure~\ref{fig4} shows that the Stark effect cannot be
suppressed, as the tip would have to be retracted well beyond the lowest tunneling
currents experimentally accessible, which are typically in the pA-range. For
Ag(111) the Stark shift results then, at the best, in a $\sim 4-5\%$ error in the
evaluation of $E_{0}$.

We now turn to the $I\agt 2$ $\mu A$ region where our model calculation no longer
reproduces the experimental data. The comparison between the two indicates that
the actual electric field in the junction is stronger than the one predicted in
our model calculation. Since in the model the tip and surface morphologies are
assumed to be constant at all tunneling currents, we hint that in the $I\agt 2$
$\mu A$ range, i.e. $R\alt 50$ k$\Omega$, the tip and the surface deform to yield
a stronger electric field. However, this deformation is reversible, since the
spectra acquired at $I=1$ nA prior and after ramping the current to $I\agt 2$ $\mu
A$ (three-step procedure described earlier) did not yield any substantial
differences. This picture agrees with calculations performed for Au(100)
\cite{Besenbacher}, which indicate that the tip and the surface undergo an elastic
deformation, i.e.\ they stretch towards each other, because of attractive adhesive
forces acting at $R \alt 100$ k$\Omega$. In particular, this is predicted to
result in a deviation from the exponential behavior of Eq.\ \ref{tunnel}, which we
do indeed observe in the $I$ versus displacement measurements at $I\agt 2$ $\mu
A$.

To summarize, in STS a downward shift of $E_{0}$ of the Ag(111) Shockley surface
state occurs when decreasing tip-surface distance. A model calculation explains
this observation in terms of a Stark shift produced by the bias and by the
lowering of the image potential compared to an isolated surface. We extract a
non-perturbed value of $-64$ meV for the binding energy of the Ag(111) surface
state. In light of the agreement with the PES measurements for Ag(111), we
conclude that a possible positive shift of $E_{0}$ in the PES data of
Ref.~\onlinecite{Reinert} due to surface defects is negligible. Finally, for
tunneling resistances $R \alt 50$ k$\Omega$, an enhanced shift is observed, which
we assign to an elastic deformation of the tip-surface morphology. These results,
supported by our recent observation of a Stark effect for other noble metal
surface states and for the Ag/Ag(111) adatom \cite{Kroeger}, suggest that STS data
require a thorough quantification of the Stark effect when striving for high
energy resolution, especially for states whose wavefunctions have large decay
lengths into vacuum.

We gratefully acknowledge G. Hoffmann whose results for Na/Cu($111$) motivated
this work, and J. Kr{\"o}ger, J. Kuntze, and S. Crampin for fruitful discussions.
L.L., T.M. and R.B. thank the Deutsche Forschungsgemeinschaft for financial
support, and P.J. the Swedish Natural Science Research Council (VR).


\begin{thebibliography}{99}

\bibitem{Crommie Hasegawa} L. C. Davis, M. P. Everson, R. C. Jaklevic, and W. Shen, \prb\textbf{43}, 3821 (1991);
Y. Hasegawa and Ph. Avouris, \prl\textbf{71}, 1071 (1993);
M. F. Crommie, C. P. Lutz, and D. M. Eigler, Nature \textbf{363}, 524 (1993).

\bibitem{ssinteractions} J. Repp, F. Moresco, G. Meyer, K.-H. Rieder, P. Hyldgaard,
and M. Persson, \prl \textbf{85}, 2981 (2000);
N. Knorr, H. Brune, M. Epple, A. Hirstein, M. A. Schneider, and K. Kern, \prb \textbf{65}, 115420 (2002).

\bibitem{ssconfinement} E. J. Heller, M. F. Crommie, C. P. Lutz, and D. M. Eigler, Nature \textbf{369}, 464 (1994);
J. Li, W.-D. Schneider, R. Berndt, and S. Crampin, \prl \textbf{80}, 3332 (1998);
K.-F. Braun and K.-H. Rieder, \prl \textbf{88}, 96801 (2002).

\bibitem{Joerg} J. Kliewer, R. Berndt, E. V. Chulkov, V. M. Silkin, P. M.
Echenique, and S. Crampin, Science \textbf{288}, 1399 (2000).

\bibitem{LifetimeB} J. Li, W.-D. Schneider, R. Berndt, O. R. Bryant, and S. Crampin, \prl \textbf{81}, 4464 (1998).

\bibitem{LifetimeK} L. B\"urgi, O. Jeandupeux, H. Brune, and K. Kern, \prl
\textbf{82}, 4516 (1999).

\bibitem{Reinert} F. Reinert, G. Nicolay, S. Schmidt, D. Ehm, and S. H\"ufner,
\prb \textbf{63}, 115415 (2001).

\bibitem{PESDefects} O. S\'anchez, J. M. Garc\'{\i}a, P. Segovia, J. Alvarez, A. L.
V\'azquez de Parga, J. E. Ortega, M. Prietsch, and R. Miranda, \prb \textbf{52},
7894 (1995);
F. Theilmann, R. Matzdorf, G. Meister, and A. Goldmann, \prb \textbf{56}, 3632 (1997).

\bibitem{Binnig} R. S. Becker, J. A. Golovchenko, and B. S. Swartzentruber, \prl \textbf{55}, 987 (1985);
G. Binnig, K. H. Frank, H. Fuchs, N. Garcia, B. Reihl, H. Rohrer, F. Salvan, and
A. R. Williams , \textit{ibid.} \textbf{55}, 991 (1985).

\bibitem{BandBending} M. McEllistrem, G. Haase, D. Chen, and R. J. Hamers, \prl \textbf{70}, 2471 (1993).

\bibitem{Besenbacher} L. Olesen, M. Brandbyge, M. R. S{\o}rensen, K. W. Jacobsen,
E. L{\ae}gsgaard, I. Stensgaard, and F. Besenbacher, \prl \textbf{76}, 1485
(1996).

\bibitem{Krans} J. M. Krans, C. J. Muller, I. K. Yanson, Th. C. M. Govaert, R.
Hesper, and J. M. van Ruitenbeek, \prb \textbf{48}, 14721 (1993).

\bibitem{Chulkov} E. V. Chulkov, V. M. Silkin, and P. M. Echenique, Surf. Sci.
\textbf{437}, 330 (1999).

\bibitem{EF} The Fermi energy is fixed to $E_{F}=5.085$ eV so that the calculated
curve with $\phi_{\rm samp}=\phi_{\rm tip}=4.56$ eV matches the experimental data.
This procedure we employ to determine $E_{F}$ differs from the approach of
Ref.~\onlinecite{Chulkov}.

\bibitem{Kroeger} J. Kr{\"o}ger, L. Limot, P. Johansson, and R. Berndt,
unpublished.

\end{thebibliography}
\end{document}